# BEAM-BASED HOM STUDY IN THIRD HARMONIC SC CAVITIES FOR BEAM ALIGNMENT AT FLASH*

P. Zhang[†‡#], N. Baboi[#], H. Ecklebe[°], T. Flisgen[°], H.W. Glock[°],
R.M. Jones[†‡], B. Lorbeer[#], I.R.R. Shinton[†‡]

[†]School of Physics and Astronomy, The University of Manchester, Manchester, U.K.
[‡]The Cockcroft Institute of Accelerator Science and Technology, Daresbury, U.K.
[#]Deutsches Elektronen-Synchrotron (DESY), Hamburg, Germany
[°]Universität Rostock, Rostock, Germany

## Abstract

An electron beam entering an accelerator cavity excites higher order modes (HOM). These are radiated to HOM couplers and subsequently damped. They can also be used to facilitate beam position monitoring. The modes excited by off-axis beams are the focus of this study. Results are presented on the first analysis of HOM dependency on the beam position for the third harmonic cavities at FLASH. The analysis is based on the beampipe and cavity dipole modes. Various analysis methods are used and compared to satisfy the needs emerging from the characteristics of those modes. The electric center of each mode is observed by looking for the minima of the HOM signals of different beam positions. A single electron bunch per RF pulse is used.

## INTRODUCTION

In FLASH [1], a cryo-module named ACC39 containing four third harmonic superconducting cavities (see Fig. 1) operating at 3.9 GHz was installed to linearize the energy spread along the electron bunch [2].

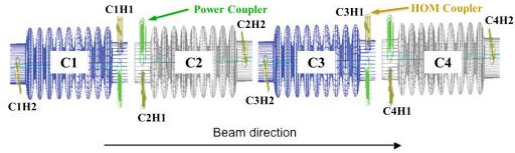

Figure 1: Schematic representation of the four cavities within ACC39 module. The power couplers are indicated in green, and the HOM couplers are indicated in yellow.

The wakefields in the third harmonic cavities are much larger than those in the main accelerating cavities (SC TESLA cavities, made of nine cells and operating at 1.3 GHz) according to the scaling relations [3],

$$W_{//} \sim \lambda^{-2}, \quad W_\perp \sim \lambda^{-3} \qquad (1)$$

where $W_{//}$ and $W_\perp$ are longitudinal and transverse wake potentials respectively, and $\lambda$ is the structure scaling factor. Besides that, the HOM spectrum is much more complex than those in 1.3 GHz accelerating cavities since larger-diameter beam pipes allow most HOMs to propagate to adjacent cavities. It is important to ensure that the HOMs are well-suppressed and their effect on the beam is minimized by aligning the beam to the electric center of the cavity. On the other hand, the beam-excited fields also enable the beam position within the cavity to be remotely determined [4][5][6]. For this purpose we plan the design of special electronics. HOM measurements both with and without beam-excitations have been performed beforehand to understand their behaviours [7][8].

Among HOMs, dipole modes have a linear dependence on beam offset [3],

$$(Amplitude)_{dipole} \sim W_\perp \sim \frac{r}{a} \qquad (2)$$

where $r$ is the beam offset and $a$ is the iris radius. The following discussion will be focused on dipole modes of the third harmonic cavities.

## HOM MEASUREMENT WITH BEAM-EXCITATIONS

### Measurement Setup

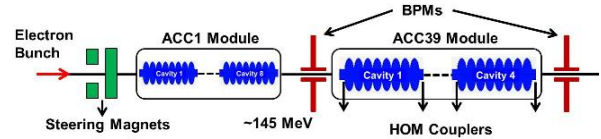

Figure 2: Schematic of beam-based HOM measurement setup (not to scale, cavities in ACC1 module are three times larger than those in ACC39 module).

The schematic of HOM measurement setup is shown in Fig. 2. An electron bunch (about 0.5 nC) has been accelerated by ACC1 before entering the ACC39 module. Two steering magnets located upstream of ACC39 can deflect the beam horizontally and vertically respectively, therefore are used to produce a transverse offset of the electron bunch in ACC39 module. Two beam position monitors (BPM) are used to monitor transverse beam positions before and after the module. By switching off the accelerating field of ACC39 and all quadruples close to ACC39, a straight line trajectory of the electron bunch is produced between those two BPMs. Therefore, the offset of electron bunch in each cavity can be determined by interpolating two BPM readouts.

### Measurement Results

Single electron bunch per RF pulse excites wakefield in the cavities. This wakefield is measured from each of the eight HOM couplers by a Tektronix Oscilloscope (Scope) of up to 6 GHz bandwidth and a Tektronix Real-time Spectrum Analyzer (RSA). A 10 dB external attenuator is attached to each HOM coupler to reduce the power of beam-excited signal radiated to the couple. The time domain signal is sampled in 20 Giga-samples/s and 200,000 points are recorded in a time window of 10 µs. A typical recorded waveform is shown in Fig. 3a. Each waveform is excited by a single electron bunch. The Scope is triggered synchronous with the beam pulse. The same trigger is also used for spectrum measurement by RSA. The HOM spectrum ranging from 4.0 to 5.5 GHz is recorded from each HOM coupler (shown in blue curve in Fig. 3b). Each 50 MHz of the spectrum is excited by a single electron bunch. A frequency step of 10 kHz and a resolution bandwidth of 22.5 kHz are used. After a FFT (Fast Fourier Transform) applied on the time-domain waveform (Fig. 3a), HOM signals from the two different devices are compared and present very good consistency, which is shown in Fig. 3b.

---

*Work supported in part by European Commission under the FP7 Research Infrastructures grant agreement No.227579.

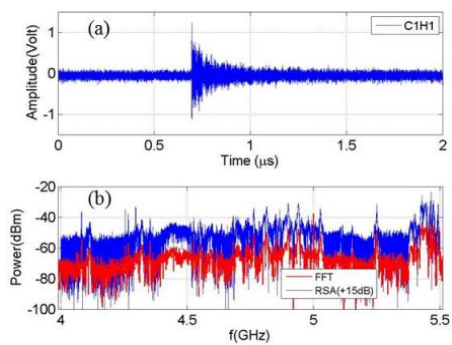

Figure 3: HOM signals from coupler C1H1. (a) Time-domain waveform; (b) Frequency-domain spectra from RSA (blue) and Scope (after FFT, red).

Besides HOM signals, the beam information is also recorded synchronously by reading nearby toroids, BPMs and currents of steering magnets. By using the steering magnets, the electron bunch is moved in various ways: two-dimensional cross movement and grid movement. The readings of the upstream BPM are shown in Fig. 4. One notices some tilt of the position readings. This is partially due to coupling between the $x$ and $y$ planes caused by the ACC1 module (shown in Fig. 2) and partially by the upstream BPM. The 2D grid data have been split into calibration samples (denoted by blue dots in Fig. 4b) and validation samples (denoted by red asterisks in Fig. 4b) for further analysis, which will be described in the later section.

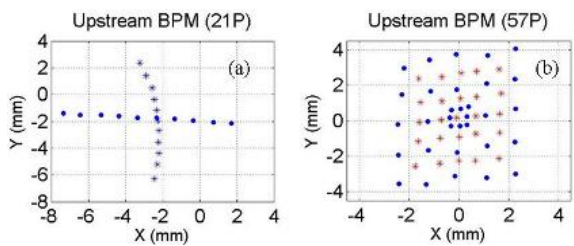

Figure 4: BPM readings during various beam movements.

## HOM DEPENDENCE ON BEAM OFFSET

### Beampipe Modes

The HOM signals of 10 different horizontal beam positions (blue dots in Fig. 4a) from 4.116 to 4.1186 GHz are shown in Fig. 5. According to simulations [9][10][11], those are most likely beampipe modes (the frequencies are below the first dipole cavity passband). The vertical position of the beam is -1.03 mm. The variation of the mode amplitudes with respect to the beam positions can be observed. The amplitude does not reach zero in this case because of contributions from the other polarization and probably of an angle in the beam trajectory.

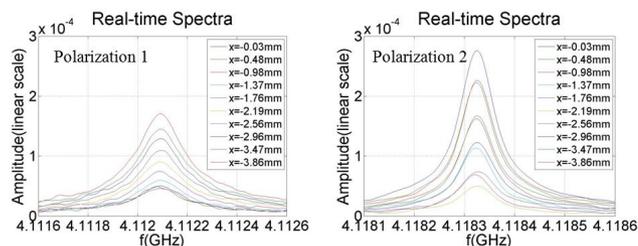

Figure 5: Frequency-domain signals of the beampipe modes from the first dipole passband measured from HOM coupler C2H2.

By Lorentzian fit, one can get the amplitude of the modes for each spectrum. Fig. 6 shows the amplitude of the two modes (Fig. 5) in linear scale as a function of the beam position interpolated in the cavity for both horizontal and vertical beam movements (Fig. 4a). The two polarizations have different electric centres, since the scan has an offset.

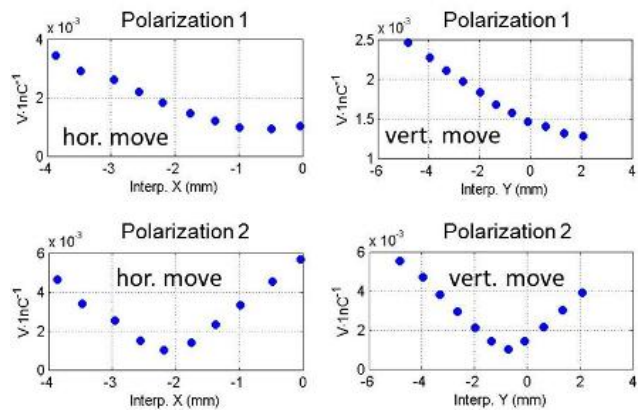

Figure 6: HOM signal amplitudes for each polarization of the beampipe mode of cavity 2 as a function of the beam position interpolated in the cavity.

Dipole mode has two polarizations perpendicular to each other, which can be seen clearly in Fig. 7 when plotting the amplitude of each mode in the 2D grid movement (Fig. 4b).

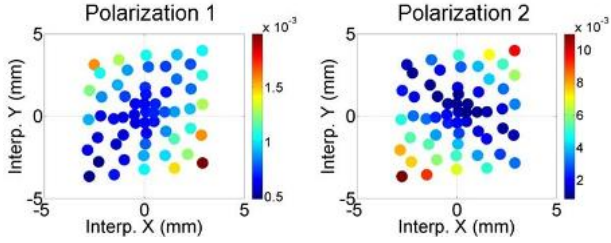

Figure 7: Mode Amplitudes of two polarizations of the beampipe mode of cavity 2.

*Cavity Modes*

Previous measurements [7][8] have shown the complexity of the spectrum of the third harmonic cavities. Most of the dipole modes are able to travel from cavity to cavity (see Fig. 3), which makes the identification of individual mode less feasible. To extract beam position information from the HOM measurements, a straightforward method is direct linear regression (DLR),

$$A \cdot M + B_0 = B \quad (3)$$

where $A$ is the matrix of HOM spectra, each row of $A$ denotes a spectrum taken from one beam movement, $B$ is the matrix of beam positions in $x$ and $y$ interpolated from two BPM readouts, each row of $B$ denotes one beam movement, $B_0$ is an offset matrix. A least-squares fit is performed to find matrix $M$ in the linear system composed by $A$ and $B$ (with an offset $B_0$).

For the 2D grid movement, the calibration samples are used to determine the matrix $M$, while the validation samples are used to compare the measured $B$ to the predicted $B$ from the determined $M$.

Fig. 8 shows the results of applying DLR to the second dipole passband spectra (5.1-5.5 GHz, measured from HOM coupler C3H2). The prediction and measurement fit very well for transverse beam offset $x$ and $y$. To measure the consistency of two curves, a coefficient $r^2$ [12] has been calculated.

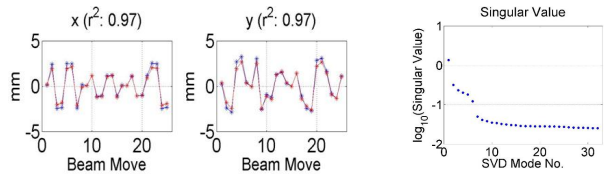

Figure 8: HOM response of the second dipole passband using DLR. Measurement (blue) and prediction (red) are compared.

Figure 9: Singular values of SVD vectors decomposed from the training samples.

As the size of matrix $M$ depends on the size of matrix $A$, the computation intensity grows dramatically with the sampling points of the spectrum. An alternative approach is to perform singular value decomposition (SVD) [13][6] on the matrix. The method looks for the main directions or patterns of the matrix in terms of SVD vectors. SVD decompose the matrix $A$ into the product of three matrices,

$$A = U \cdot S \cdot V^T \quad (4)$$

where $U$ and $V$ are unitary matrices, $S$ is the matrix consisting of singular values and $V^T$ indicates the transpose of $V$. Based on singular values of each SVD vector (shown in Fig. 9), the first 6 SVD vectors have been selected to create a matrix of SVD vector amplitude, then regressed against the matrix $B$ by using Eq. 3. The results on training samples show how representative the first 6 SVD vectors are to describe the entire spectrum (see Fig. 10a), and Fig. 10b shows the prediction power of the first 6 SVD vectors.

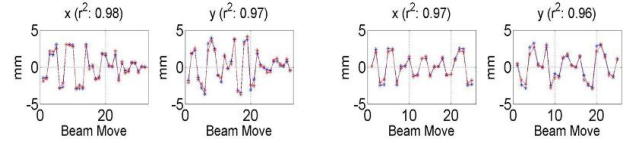

(a) Calibration samples    (b) Validation samples

Figure 10: HOM response of second dipole band using SVD. Measurement (blue) and prediction (red) are compared.

## CONCLUSIONS

HOM dependence on beam movement has been observed for the first time in third harmonic cavities at FLASH. There are beampipe modes which do not propagate through adjacent cavities; that enables the beam position to be determined for each individual beampipe. Those modes are identified and the polarizations of the modes are observed. On the contrary, cavity modes of the first two dipole passbands propagate; that makes the spectrum complex enough to drop mode identification method. Alternatively, DLR and SVD methods are used to abstract beam position information from the HOM spectra. Both methods show robust prediction power. Future work includes further investigation of suitable higher order modes (the fifth dipole passband is currently being investigated) for beam and cavity alignment monitoring. The time-domain signals are also being studied for the HOM electronics.